\documentclass[aps,prb,twocolumn,superscriptaddress,showpacs]{revtex4}
\usepackage{hyperref}
\usepackage[dvips]{graphicx}
\usepackage{amssymb,amsmath,amsfonts}
\usepackage[T1]{fontenc}
\newcommand{\alfa}{$\alpha_{\mathrm{hex}}$}
\newcommand{\dint}{$d_{\mathrm{int}}$}
\newcommand{\doping}{$\Delta E_{F}$}

\begin{document}
% Use the \preprint command to place your local institutional report
% number in the upper righthand corner of the title page in preprint mode.
% Multiple \preprint commands are allowed.
% Use the 'preprintnumbers' class option to override journal defaults
% to display numbers if necessary
%\preprint{}

%Title of paper
\title{Doping of graphene by a Au(111) substrate: Calculation strategy within the local density approximation and a semiempirical van der Waals approach}

%%%%%%%%%%%%%%%%%%%%%%%%%%%%%%%%%%%%%%%%%%%%%%%%%%%%%%%%%%%%%%%%%%%%%
%% The document title should be given as usual. Some journals require
%% a running title from the author: this should be supplied as an
%% optional argument to \title.
%%%%%%%%%%%%%%%%%%%%%%%%%%%%%%%%%%%%%%%%%%%%%%%%%%%%%%%%%%%%%%%%%%%%%

\author{J. S\l awi\'{n}ska}
%\email{jagoda.slawinska@uni.lodz.pl}
\affiliation{Theoretical Physics Department II, University of Lodz, Pomorska 149/153, 90-236 Lodz, Poland}
\affiliation{Solid State Physics Department,University of Lodz, Pomorska 149/153, 90-236 Lodz, Poland}
\author{P. Dabrowski}
\affiliation{Solid State Physics Department,University of Lodz, Pomorska 149/153, 90-236 Lodz, Poland}
\author{I. Zasada}
\affiliation{Solid State Physics Department,University of Lodz, Pomorska 149/153, 90-236 Lodz, Poland}

%\email[]{Your e-mail address}
%\homepage[]{Your web page}
%\thanks{}
%\altaffiliation{}

%Collaboration name if desired (requires use of superscriptaddress
%option in \documentclass). \noaffiliation is required (may also be
%used with the \author command).
%\collaboration can be followed by \email, \homepage, \thanks as well.
%\collaboration{}
%\noaffiliation
\vspace{.15in}

\begin{abstract}
We have performed a density functional study of graphene adsorbed on Au(111) surface using both a local density approximation and a semiempirical van der Waals approach proposed by Grimme, known as the DFT-D2 method. Graphene physisorbed on metal has the linear dispersion preserved in the band-structure, but the Fermi level of the system is shifted with respect to the conical points which results in a doping effect. We show that the type and amount of doping depends not only on the choice of the exchange-correlation functional used in the calculations, but also on the supercell geometry that models the physical system. We analyzed how the factors such as the in-plane cell parameter and interlayer spacing in gold influence the Fermi level shift and we found that even a small variation in these parameters may cause a transition from p-type to n-type doping. We have selected a reasonable set of model parameters and obtained that graphene is either undoped or at most slightly p-type doped on the clean Au(111) surface, which seems to be in line with experimental findings. On the other hand, modifications of the substrate lattice may induce larger doping up to 0.30-0.40 eV depending on the graphene-metal adsorption distance. The sensitivity of the graphene-gold interface to the structural parameters may allow to tune doping across the samples which could lead to possible applications in graphene-based electronic devices. We believe that the present remarks can be also useful for other studies based on the periodic DFT.
\end{abstract}

\maketitle

% body of paper here - Use proper section commands
% References should be done using the \cite, \ref, and \label commands

\section{Introduction}
Graphene, a two-dimensional sheet of carbon atoms packed into a honeycomb lattice, has attracted considerable attention due to its thin geometry and unusual electronic properties promising for a wide range of applications.\cite{novoselov_science, review, nanoletters} One of the most important themes in recent research is to investigate the interaction between graphene and its surrounding environment, which is of practical relevance for the process of samples production and fabrication of graphene-based electronic devices. In particular, metallic substrates are used as catalysts in graphene formation, as probes during the electrical measurements and as source/drain electrodes in electronic devices. 

Early systematic density functional theoretical studies based on the local density approximation (LDA) have shown that two types of interfaces can be formed between graphene and metal, i.e. physisorption and chemisorption should be distinguished.\cite{holendrzy_prl, holendrzy} The bonding of graphene to Al, Ag, Cu, Au and Pt(111) surfaces is weak and preserves graphene's characteristic Dirac cones, while Co, Ni and Pd bind graphene strongly which disturbs its $\pi$ bands through coupling with the d-orbitals of the metals. The physisorption of graphene on the metal surface causes a Fermi level shift downward (upward): It means that holes (electrons) are donated by the metal substrate to graphene which becomes p-type (n-type) doped. According to this theoretical study, graphene on Al, Ag and Cu is n-type doped, while the interaction with Au and Pt (111) surfaces causes p-type doping. Most of these LDA results were confirmed in experiments.\cite{cobalt, nikiel2, pallad, sers, klusek, platyna}

On the other hand, the local density approximation is well known to overestimate the binding in weakly bonded systems where interactions are mainly of van der Waals (vdW) type. It is regarded to be an unreliable method for inhomogeneous systems, such as graphite and many organic compounds\cite{vanin}. A generalized gradient approximation (GGA) does not improve the results. It tends to underestimate the binding: in the case of graphene on gold it gives no binding at all, while for graphene on Ni only a metastable minimum is predicted. Although GGA in general offers reasonable predictions for metals, it gives incorrect binding behavior for the graphene-metal\cite{miguel, vanin} and for other organic-inorganic interfaces\cite{interface_xu, pentacen}. Thus, much research effort has been devoted to include vdW forces in the studies of graphene-metal composites.

In the paper of Vanin \textit{et al.}\cite{vanin} the recently developed van der Waals density functional (vdW-DF)\cite{dion1, dion2} has been used to study the metal-graphene interfaces. Contrary to the previous LDA results, the inclusion of non-local correlations leads to weak binding for all metals and gives graphene-substrate separations in the range of 3.40-3.72 \AA. At these distances graphene is physisorbed and the graphene band structure is unaffected by the substrate: Strong binding in the case of Ni, Co and Pd is not predicted by the vdW-DF, thus there is no band-gap opening at the K-point. These results are in a strong conflict with low-energy electron diffraction (LEED)\cite{nikiel} and angle-resolved photoemission spectroscopy (ARPES) measurements\cite{nikiel2} for graphene on Ni, which predict an adsorption distance of 2.11 \AA\, and the presence of a band-gap. It seems that the DFT calculations should be tested first for the graphene-Ni interface so that the adsorption distance and band structure are correct.

Furthermore, in a very recent study\cite{comparative} of graphene adsorbed on (111) surfaces of Ni, Cu, Pd, Ag, Au and Pt the second version of vdW-DF (vdW-DF2)\cite{lee} with a C09 exchange functional developed by Cooper\cite{c09} has been applied and assessed to be the most appropriate combination of exchange-correlation interaction. It gives accurate adsorption geometries and electronic structures in accordance with available experimental data for most of the metals. Two classes of interfaces, physisorption and chemisorption, are distinguished, in agreement with previous LDA results. The band-gap opening at the K point observed on the Ni (111) is reproduced reasonably well. The authors also compare their graphene-copper distance with semiempirical DFT-D (Ref.\onlinecite{grimme}) calculations and conclude that both methods give nearly the same equilibrium distances.

One important discrepancy when comparing to previous theoretical and experimental studies, is reported in Ref.\onlinecite{comparative}. The obtained n-type doping of graphene on gold contradicts the experimental findings of Klusek \textit{et al.}, who observed that holes are donated to graphene on Au substrate. The p-type doping has been confirmed in the high-resolution ARPES measurements of Varykhalov \textit{et al.}\cite{arpes}. It has been demonstrated that graphene on Ni intercalated by a Au monolayer remains gapless and that a very small hole doping ($\Delta E_{F}=+100\pm20$ meV) is observed in the band structure.
Moreover, it has been shown that also the presence of gold atoms on graphene causes its p-type doping\cite{gierz, sers}. In the Ref.\onlinecite{comparative}, it is suggested that the origin of this inconsistency for the graphene-Au system is connected with the sensitivity of the doping polarity to the adsorption distance. It seems that a more accurate description of the graphene-gold contact is needed.

Theoretical research concerning graphene on the metals' surfaces has been focused on the construction and choice of the appropriate exchange-correlation functionals. They should provide a correct determination of the graphene-substrate separation and, as a consequence, capture the main characteristic features of the spectrum, as an energy gap in the band structure of chemisorbed graphene as well as the type and amount of doping for weak adsorption. The correct determination of the adsorption distance is especially crucial, when the transition between extremely different properties occurs near the estimated distance, in particular the change from n-type to p-type doping of graphene on gold\cite{holendrzy}.

However, there are other factors that are important in the modeling of interfaces, especially those related to the geometry of the supercell. First, performing periodic DFT calculations requires direct matching of the graphene's and substrate's unit cells. Typically, the C-C bonds of physisorbed graphene are not stretched or quenched in the physical sense. On the other hand, adjusting the substrate to the graphene generates strain in its lattice, because the interatomic distances in a metal are artificially contracted. It results in a change in the surface relaxation and can additionally alter the electronic properties. In the case of a slight lattice mismatch, these two competing geometries are considered equivalently in literature\cite{interface_xu, holendrzy}. Furthermore, there are various routes for the structural optimization of the graphene-metal system. Relaxation leads to different geometries and electronic properties if i) the metal atoms are kept in their experimental posistions\cite{vanin}, ii) only carbon atoms are fixed\cite{gong}, or iii) graphene and the two top layers of gold are allowed to move\cite{holendrzy}. The above-mentioned factors give small percentage changes in results for almost all types of the substrates. However, the  graphene-gold interface is very sensitive to variations in the adsorption parameters and qualitatively different properties can be derived.

In this paper, we attempt to resolve the issue of doping graphene on Au(111). In our DFT calculations both local density and generalized gradient approximations have been used. To include vdW corrections, we have applied the semi-empirical DFT-D2 Grimme's method\cite{grimme, bucko}, which has brought satisfactory predictions for similar systems. For example, the adsorption of aromatic molecules on the (111) surfaces of noble metals\cite{aromatic} and the adsorption of Cu, Ag and Au atoms on graphene\cite{adsorption} were accurately described. DFT-D2 should be then a valuable tool also for graphene-metal contact studies.

First, we have performed a series of fixed geometry calculations to clearly identify the main model factors determining the doping of graphene on gold. We have calculated the total energies as well as the shifts of the Fermi level with respect to the Dirac points using two different functionals. We show that the choice of exchange-correlation interaction approximation alters the doping directly and also indirectly by influencing the adsorption distance. Second, we have evaluated the two-dimensional maps of Fermi level shifts as a function of graphene-gold separations and the in-plane unit cell parameters assumed for the simulations in order to identify regions where graphene is n-type doped, p-type doped or undoped. We have analyzed how the Fermi level shift varies with a change of interlayer spacings in gold, which have also significant physical implications. Finally, we have calculated surface relaxation and studied its influence on electronic properties. This detailed analysis allows to assign the inconsistency indicated in Ref. \onlinecite{comparative} to geometry parameters defined in the model. 

The main purpose of this paper is to develop a methodology, so that the calculations with vdW corrections give correct results for all typical metals and to propose a set of model parameters which are reasonable from both computational and experimental points of view. Since the recommended choice of interface geometry prevents a reliable relaxation of the Au(111) surface, alternative routes to manage this limitation are suggested. Moreover, in order to demonstrate the reliability of a simple DFT-D2 method, we have performed basic calculations of a graphene-nickel interface.

The present paper is organized as follows. In Sec. II the calculation method and computational details are summarized. In Sec. III we present the interpretation of fixed-geometry calculations and identify factors determining the doping. Surface relaxation is analyzed in Sec. IV, while conclusions and perspectives are discussed in Sec. V.

\section{Methods}
\begin{table*}
\caption{\label{tab:table3}The sets of parameters determining the size of the supercell and the geometry of the slab. For bulk Au the LDA-optimized lattice constant compare well with the experimental value\cite{kittel}. The one obtained using the PBE functional is overestimated, but in accordance with previous DFT results\cite{previous}. The interlayer spacings in gold $d_{\mathrm{int}}$ are calculated from Eq. (2). All values are given in \AA.}
\begin{ruledtabular}
\begin{tabular}{ccccccccccccc}
 &\multicolumn{3}{c}{LDA}&\multicolumn{3}{c}{PBE}&\multicolumn{3}{c}{PBE+D2}&\multicolumn{3}{c}{experimental}\\
 \,&$a$ &$\alpha_{\mathrm{hex}}$ &$d_{\mathrm{int}}$ &$a$ 
&$\alpha_{\mathrm{hex}}$ &$d_{\mathrm{int}}$ &$a$ &$\alpha_{\mathrm{hex}}$ &$d_{\mathrm{int}}$ &$a$ &$\alpha_{\mathrm{hex}}$ &$d_{\mathrm{int}}$\\ \hline
 Graphene&2.445&4.89&2.305&2.468&4.936&2.327&2.466&4.932&2.325&2.46&4.92&2.320 \\
 Gold&4.063&4.976&2.345
 &4.173&5.111&2.41&4.102&5.024&2.368&4.08&4.997&2.356\\
\end{tabular}
\end{ruledtabular}
\end{table*}
\begin{figure}
\includegraphics[width=0.45\textwidth]{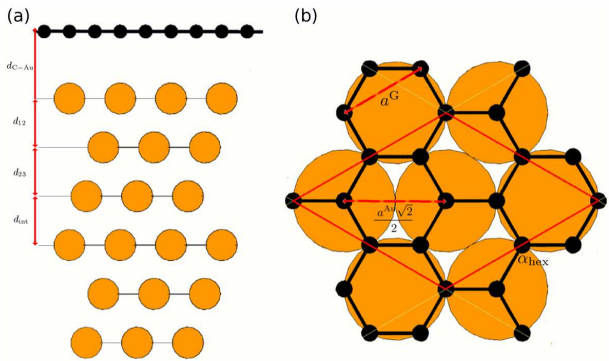}
\caption{\label{geometria} (Color online) Side view (a) and top view (b) of adsorption geometry of graphene on Au(111). Carbon atoms are denoted as black balls, gold atoms as yellow (gray) ones. Parallelogram defines the unit cell.}
\end{figure}

We have performed \textit{ab initio} DFT calculations using the \textsc{vasp} code\cite{vasp1, vasp2} equipped with the projector augmented wave (PAW) method\cite{paw1, paw2} for electron-ion interactions. The exchange-correlation interaction is treated in the generalized gradient approximation in the parametrization of Perdew, Burke and Ernzerhof (PBE)\cite{pbe}. Long-range dispersion corrections have been taken into account within a DFT-D2 approach of Grimme\cite{grimme}, as implemented in the latest version of \textsc{vasp}\cite{bucko}. Since a dispersion coefficient $C_{6}$ for gold has not been listed in the original paper of Grimme\cite{grimme}, we have used a value of 40.62 Jnm$^{6}$/mol and of 1.772 \AA\, for the vdW radius of Au ($R_{0}$), according to the suggestion given in Ref. \onlinecite{adsorption}. Also, as in Ref. \onlinecite{adsorption}, the pair interactions up to a radius of 12 \AA\, have been included in the calculations and the global scaling factor $s_{6}$ has been set to 0.75 because the PBE functional was chosen. For comparison, most of the calculations have been also done using the local density approximation (LDA)\cite{perdewzunger}.  The electronic wave functions have been expanded in a plane-wave basis set of 400 eV. The electronic self-consistency criterion has been set to $10^{-7}$ eV. 

The Au(111) surface is modeled by a periodic slab geometry: each supercell contains six atomic layers of metal and a graphene sheet adsorbed on one side. A vacuum spacing of at least 13\, \AA\, is used in the direction normal to the interface in order to avoid interactions with spurious replicas. We employed the in-plane adsorption geometry suggested in Refs.\,\onlinecite{holendrzy_prl, holendrzy} which is illustrated in Fig.\ref{geometria}, where 2$\times$2 graphene and substrate  ($\sqrt{3}\times\sqrt{3}$) unit cells are directly matched. We have performed calculations for both configurations, i.e. with the lattice constant of the metal adapted to that of graphene and with the graphene lattice constant adjusted to the substrate one. The in-plane cell parameter $\alpha_{\mathrm{hex}}$ can be defined as:
\begin{equation}
\alpha_{\mathrm{hex}}^{\mathrm{G}}=2a^{\mathrm{G}}, \quad \alpha_{\mathrm{hex}}^{\mathrm{Au}}=a^{\mathrm{Au}}\sqrt{\frac{3}{2}}
\end{equation}
%if graphene is adjusted to the substrate, and
%\begin{equation}
%\alpha_{\mathrm{hex}}=2a
%\end{equation}
%if gold is contracted to graphene. The parameter 
where $a^{\mathrm{G}}$ and $a^{\mathrm{Au}}$ stand for the lattice constant of graphene and bulk gold, respectively.

Consistently, the Au(111) interplane distances can be calculated from:
\begin{equation}
d_{\mathrm{int}}=\frac{\alpha_{\mathrm{hex}}\sqrt{2}}{3}
\end{equation}
However, this can give too short distances if the metal is adjusted to the graphene, and it seems to be more reasonable to set it independently, especially when relaxation is not included. The above-mentioned parameters as well as their relation to the lattice constants are illustrated in Fig.\ref{geometria}. Note that in the case of structural optimization the parameter \dint\, is no longer a constant. Instead, there appear different values $d_{12}$, $d_{23}$, etc. We define $d$ as the fixed graphene-substrate distance and $d_{C-Au}$ as optimized distance between graphene and gold.

Moreover, graphene's and metal's lattice constants can be determined according to various approaches. They can be either optimized using LDA and PBE functionals, or measured experimentally. It gives a wide choice of values that determine the size of the supercell (\alfa) and the corresponding interlayer distances \dint. Typical sets of these parameters calculated from different values of $a^{\mathrm{G}}$ and $a^{\mathrm{Au}}$ are listed in Table I. We have performed calculations for most of these sets.

The DFT-D2 method allows to include vdW corrections within each ionic step. During the structural optimizations done with a conjugate gradient algorithm the top two layers of gold as well as all carbon atoms have been relaxed. Total energies were converged to within $10^{-6}$ with respect to the ionic steps.

For accurate Brillouin zone integrations we use the tetrahedron scheme\cite{tetrahedron} and the $\Gamma$-centered 24$\times$24$\times$1 k-point mesh. 
 
As a benchmark test of the PBE+D2 approach applied to graphene/metal interfaces, we first optimized the geometry of graphene/Ni system and determined its electronic properties. All parameters needed in calculations are the same as those  described above. (The $C_{6}$ and $R_{0}$ coefficients for Ni are defined in \textsc{vasp}.) Moreover, the spin polarization is taken into account. The supercell geometry is set as proposed in Ref.\onlinecite{holendrzy}, where the nickel ($1\times 1$) unit cell is adjusted to graphene with a LDA-optimized lattice constant equal to $a=2.445$ \AA. The LDA calculations of the structure lead to the adsorption distance of 2.05 \AA\,, thus the results given in Ref.\onlinecite{holendrzy} are well reproduced. In contrast, the PBE+D2-optimized graphene-nickel distance achieves nearly 2.15 \AA\, (2.11 without spin polarization) that is still in line with experimental findings\cite{nikiel}. A band-gap at the graphene K-point is comparable to the one obtained within local density approximation, which is smaller than the one observed in the ARPES experiment\cite{nikiel2}. The use of vdW-DF2$^{\mathrm{C}09}$ might give more accurate values, but we believe that the DFT-D2 method is sufficiently precise for a description of graphene-metal systems.
\section{Fixed geometry calculations: main factors determining the doping}
\begin{figure}
\includegraphics[width=0.9\columnwidth]{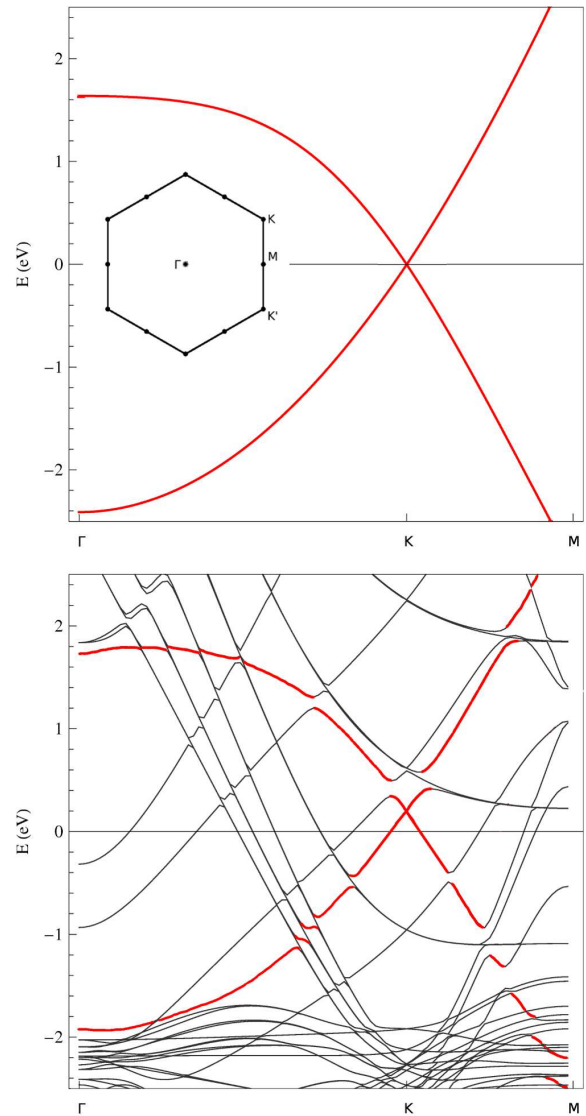}
\caption{\label{pasma} (Color online) Top panel: The electronic structure of free-standing graphene obtained with a $2\times2$ unit cell. The inset shows a Brillouin zone with the high-symmetry points labeled by $\Gamma$, K and M. Bottom panel: The band structure of graphene on Au(111). The red lines denote the graphene bands with a $\pi$ character. The zero energy is at a Fermi level, \dint=2.35 \AA,\, $a=2.445$\,\AA,\, $d=3.31$\,\AA.}
\end{figure}

The weak binding of graphene on gold preserves the characteristic conical $\pi$ bands touching at the K/K' points. In contrast to the free-standing layer, the Fermi level is shifted with respect to the Dirac points due to the interaction with the metal surface. We define the doping level as $\Delta E_{F}=E_{D}-E_{F}$, where $E_{D}$ is the energy where the Dirac cones touches and $E_{F}$ denotes a Fermi energy of the system. In Fig.\ref{pasma} we present the band-structure of free-standing graphene and of graphene on gold calculated in a fixed geometry using a local density approximation. In the latter case, the Fermi level shift \doping\, is well visible and estimated to be about 0.20 eV. The band-structure is very similar to the one presented in Fig.2 of Ref.\onlinecite{holendrzy} in spite of the fact that relaxation has not been included. Overall, the whole dispersion hardly changes with the aforementioned model parameters: only a doping level \doping\, can be altered significantly. We have then performed fixed-geometry calculations, since it allows to discuss the role of each parameter separately.

\subsection{Graphene-substrate distance}
\begin{figure}
\includegraphics[scale=1.0]{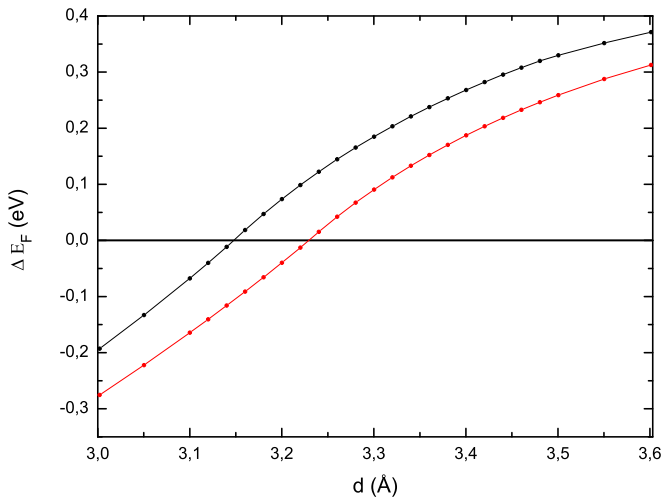}
\caption{\label{domieszkowanie} Fermi level shift with respect to the Dirac point as a function of graphene-gold distance $d$. The values of $\alpha_{\mathrm{hex}}=4.89$\,\AA\, and $d_{\mathrm{int}}=2.35$ \AA\, have been set for the calculations. The black (upper) curve has been obtained with a LDA approximation and the red one with a PBE functional.}
\end{figure}
\begin{figure}
\includegraphics[width=1.0\columnwidth]{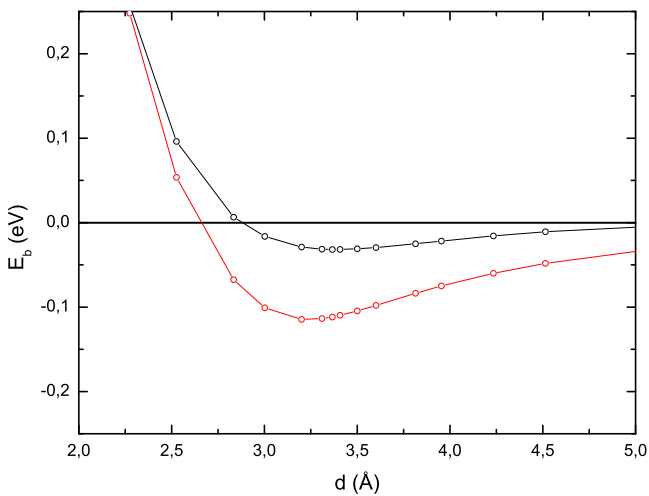}
\caption{\label{etotal} Binding energy ($E_{b}$) per carbon atom of graphene on Au(111) as a function of graphene-gold distance $d$ calculated with a LDA (upper, black line) and the PBE+D2 method (red line). The parameters are set to \alfa=4.89 \AA\, and $d=2.35$ \AA.}
\end{figure}
It has been previously noted that the doping level and polarity are very sensitive to adsorption distance. A variation in the Fermi level shift as a function of graphene-substrate separation calculated within a LDA is shown in Fig.\ref{domieszkowanie} (black upper line). Similar dependencies have been presented in Fig. 6 of Ref.\onlinecite{holendrzy} for all typical metallic substrates. One can observe\cite{holendrzy} that graphene on Al(111) is estimated to be n-type doped for a whole range of distances, while for the Pt substrate simulations consistently yield p-type doping. On the other hand, for Ag, Au and Cu the change in doping polarity from n-type to p-type occurs in the vicinity of the estimated adsorption distances. Moreover, as one can easily see in Fig.\ref{domieszkowanie} and in Ref.\onlinecite{holendrzy} (Fig. 6), such dependencies vary significantly in this region, i.e. in the vicinity of the transition the slope of the curve is significant, which explains an extreme sensitivity of doping to the graphene-metal separation. Both effects may inhibit reliable determination of the type and level of doping, especially if the distance has not been measured experimentally and its estimation relies only on DFT calculations. 

We have evaluated the same dependency using the PBE parametrization for a description of the exchange-correlation interaction (the red bottom line in Fig.\ref{domieszkowanie}). The vdW corrections are added to the conventional Kohn-Sham DFT energies, thus they do not alter the band-structure\cite{aromatic}. It should be stressed that the doping level differs from that obtained using the LDA, even if the issue of adsorption distance is ignored. The lines in Fig.\ref{domieszkowanie} illustrate that there are regions (3.15-3.23 \AA), where the LDA and PBE provide opposite doping polarities, although the numerical difference is small. A similar comparison has been presented in Fig. 9 of Ref.\onlinecite{holendrzy} for a graphene-copper system. However, the Fermi level shifts \doping\, calculated with the GGA were within ~0.07 eV of the ones obtained with the LDA, which is irrelevant if it does not lead to qualitative changes.

Obviously, the adsorption distance depends on the choice of the functional. It can be easily noticed in Fig.\ref{etotal}, where the binding energies as a function of distance have been plotted. The black (top-most) line denotes the LDA-obtained results, whereas the PBE+D2 predictions are given by the red curve. The PBE+D2 calculations lead to an adsorption distance of 3.22 \AA, while the LDA estimation is $d=3.36$\,\AA. This gives a difference in doping from zero to more than +0.25 eV (see Fig.\ref{domieszkowanie}), and thus both direct and indirect influences of the functional lead to a significant general change.

\subsection{In-plane lattice constant}

In Sec.II we have indicated that the size of the supercell parametrized by \alfa\, can be set to any value from those listed in Table I. In literature, the extreme values, 4.89 and 5.11 \AA, have been used equivalently. However, the physical meaning of this parameter is a stretch of graphene bonds, although they should be rather unaffected by weak adsorption. 

\begin{figure}
\includegraphics[width=1.0\linewidth]{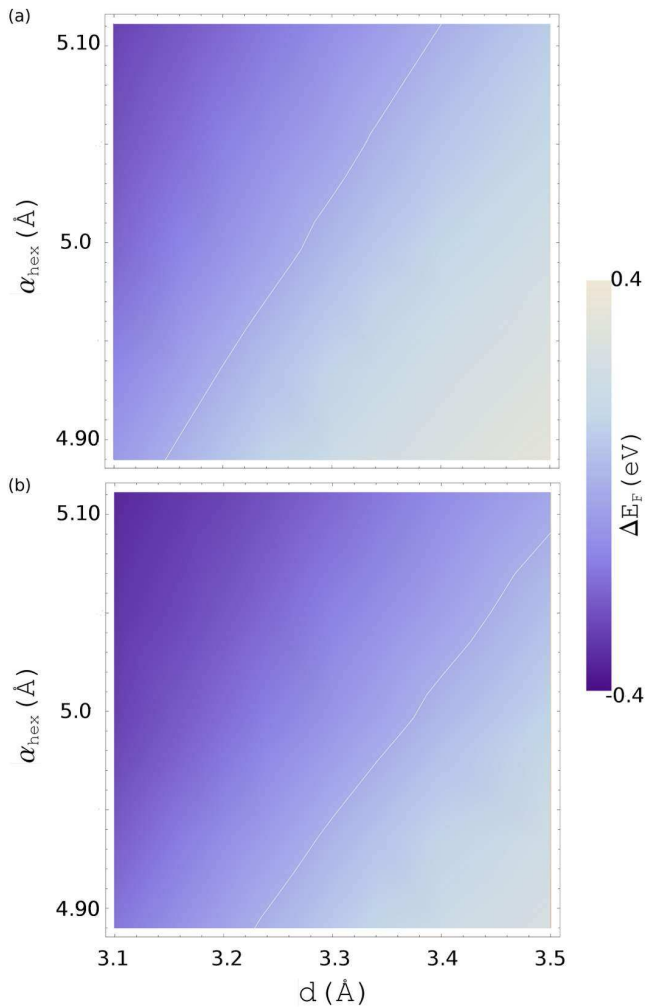}
\caption{\label{mapy}(Color online) The Fermi level shift relative to the conical point as a function of in-plane lattice constant \alfa\, and graphene-gold distance $d$. The doping changes from n-type (darker) to p-type (lighter). The white line denotes a contour of zero doping. The right-hand panel shows the color scheme used in the map. Calculations has been done using the LDA approximation and spacing \dint=2.35 \AA\, (a) and using the PBE parametrization and interplane distance \dint=2.35 \AA\, (b)}.
\end{figure}

We have calculated dependencies similar to those presented in Fig.\ref{domieszkowanie} for a few values of $\alpha_{\mathrm{hex}}$ and interpolated them into a whole range of in-plane lattice constants (4.89-5.11\,\AA). In Fig.\ref{mapy} we present the LDA-obtained two-dimensional map of Fermi level shifts $\Delta E_{F}(d, \alpha_{\mathrm{hex}})$. The color represents a value of doping for a given pair of in-plane cell parameter and graphene-gold distance ($\alpha_{\mathrm{hex}}$, $d$). One can easily see that graphene could be n-type doped if the C-C bonds were stretched sufficiently. It is clear, however, that in the case of LDA calculations setting such values of \alfa\, is, in fact, unreasonable, because LDA optimized values are \alfa=4.89 \AA\, or \alfa=4.976 \AA\, (see Table I). It is worthwhile to notice that even taking the experimental \alfa=4.997 \AA\ value, results in a considerable decrease in the value of p-type doping.

The influence of these effects has been carefully studied in Ref.\onlinecite{holendrzy} for copper substrate due to the fact that the lattice mismatch between graphene and the Cu(111) surface is the largest among the substrates considered in this study. A comparison between configurations with copper's lattice constant adapted to graphene and with graphene stretched to the in-plane lattice constant of Cu gives errors within 0.15 eV. It seems that this result could influence only the value of n-type doping, because the transition of polarities occurs for distances above the optimized one in the case of both configurations (see Fig.9 in Ref.\onlinecite{holendrzy}). However, for the gold substrate a relative shift of 0.15 eV can lead to more considerable changes or even cause a doping type transition. It seems that the lattice mismatch should not be the only criterion for such investigations.

A similar map calculated with the PBE functional is shown in Fig.\ref{mapy} (b). It can be noticed that the variation in doping is even stronger in a relevant range of $d$, which is related to the fact that the PBE dependency lies below the LDA one (Fig.\ref{domieszkowanie}). We would like to stress that for \alfa=5.11 \AA, as chosen in Ref.\onlinecite{comparative}, the electrons are always donated to graphene from the Au substrate resulting in n-type doping. We have also performed similar calculations with a slightly increased value of \dint. It yields a systematic shift of doping, but does not invalidate the overall conclusion. 

\subsection{Interlayer spacing in gold}
According to the remarks given in Sec. II, we can choose the interlayer spacing in gold \dint\, from a range of values defined in Table I. The choice of the starting \dint\, is less important if the structure is allowed to relax. Since locking the in-plane lattice constant may lead to incorrect optimization, we assume that it is better to study first the configurations with fixed \dint\, spacings chosen around the experimental bulk values.

\begin{figure}
\includegraphics[width=1.0\columnwidth]{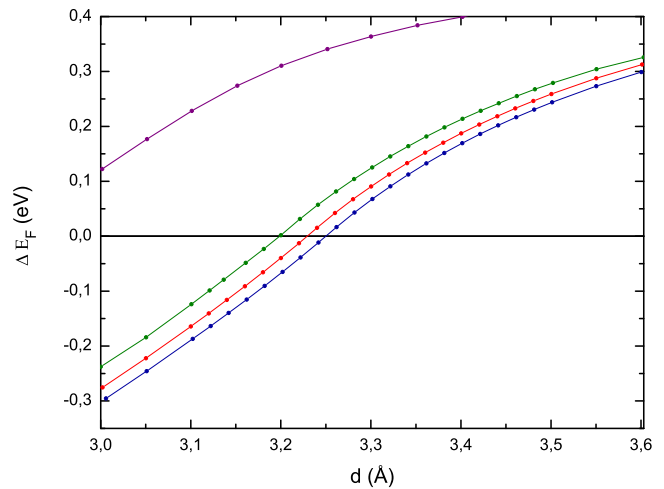}
\caption{\label{dint}(Color online) The Fermi level shift relative to the conical point as a function of graphene-metal distance for $\alpha_{\mathrm{hex}}=4.89$\,\AA\, and for different values of $d_{\mathrm{int}}$: the blue (bottom) line is assigned to $d_{\mathrm{int}}=2.305$, a red one (next) to 2.345\, \AA, the green one to 2.41\,\AA\, and the purple (the top-most) one to 2.87\,\AA. The PBE functional has been used in the calculations.}
\end{figure}

Figure \ref{dint} illustrates the dependencies $\Delta E_{F}(d)$ for different values of \dint. The lines in the middle represent typical choices for the interplane spacing (see Table I). We conclude that the larger values of \dint\, lead to an increase in the shift of the Fermi level.

The most interesting feature of the plot in Fig.\ref{dint} is the top-most (purple) line representing the function $\Delta E_{F}(d)$ for a large value of \dint. First, it allows to clearly identify the direction of the doping changes with an increase in the value of \dint. The properties of such an artificial surface also suggest how strongly the substrate modifications may affect the doping. It seems in this case that for distances $d$ higher than 3.4 \AA\, the doping saturates and achieves a value of \doping=0.40 eV. It means that any crystal imperfections of the substrate can induce changes in doping from zero to about 0.30-0.40 eV depending on the adsorption distance.

%model zamiast simulate
\section{Surface relaxation}

\begin{table*}
\caption{\label{tab:table4} Optimized structural parameters and the calculated Fermi level shifts $\Delta E_{F}$ in graphene-gold system for different starting configurations (\alfa, \dint). In the first column with the slashes, we present also the results for a clean Au surface. The fourth column with the slashes, beside standard data, contains also the results of relaxation of only carbon atoms and of a clean metal surface, respectively. All distances are given in \AA.}
\begin{ruledtabular}
\begin{tabular}{cccccccccccccccccc}
&\multicolumn{3}{c}{LDA}&\multicolumn{4}{c}{PBE+D2}\\
  \hline
 \alfa &4.89&4.976&4.89&4.89&5.02&4.89 \\
 $d_{\mathrm{int}}$ &2.345&2.345&2.41&2.345&2.345&2.41 \\
 $d_{\mathrm{C-Au}}$ &3.365/-&-&3.35&3.21/3.22/-&3.20&3.20 \\
 $d_{12}$ &2.43/2.425&2.36&2.425
 &2.52/2.345/2.55&2.43&2.54\\
 $d_{23}$ &2.39/2.39&2.34&2.39
 &2.44/2.345/2.45&2.36&2.46\\
 $\Delta E_{F}$ (eV)&0.26/-&-&0.25
 &0.08/0.00/-&-0.16&0.07\
\end{tabular}
\end{ruledtabular}
\end{table*}

DFT studies of graphene/metal interfaces typically include structural optimization. The surface relaxation not only could alter the doping level, but also is crucial for the evaluation of phonon dispersion and (scanning tunneling microscopy) STM topographies. In the previous section, we have reported the results of fixed-geometry calculations in order to analyze each model parameter separately and study its influence on the electronic properties. Now, we show that the relaxation of compressed unit cell of the substrate could lead to an unphysical expansion of its interlayer spacings. It means that performing calculations in the configuration with at least the metal atoms fixed might give more reasonable results.

The structural optimization of a system consisting of graphene and metal can be done by many ways. According to suggestion given in Ref.\onlinecite{holendrzy} we relax all carbon atoms and the top two layers of gold until the total energy change between subsequent steps is smaller than $10^{-6}$ eV. It should be mentioned that other approaches are frequently used: the relaxation of carbon atoms only\cite{vanin}, relaxation of gold atoms only\cite{gong}, relaxation of graphene and gold separately followed by distance optimization\cite{comparative} etc. Obviously, none of these methods can take into account the unique herringbone reconstruction of the clean (111) surface of Au\cite{herringbone1, herringbone2, herringbone3}. We have demonstrated that in-plane optimization hardly affects the band-structure, however the influence of reconstruction is very difficult to estimate within DFT with periodic boundary conditions and was not taken into account.

We focus on the changes in the interlayer spacing in gold upon relaxation, since we have shown that it could influence the doping level $\Delta E_{F}$. We study the relaxation effects in metal taking into account first $d_{12}$ and second $d_{23}$ interplane distance in gold. We have performed structural calculations using both LDA and PBE exchange-correlation functionals. The results for various configurations of starting geometry are listed in Table II. In particular, the relaxation of a clean Au(111) surface,  calculations with a larger unit cell, as well as optimization of only carbon atoms have been performed (see the caption of Table II). First, it should be stressed that matching gold to  graphene leads to considerable expansions of $d_{12}$ and $d_{23}$, even for clean Au(111) surface. Although the LDA and PBE+D2 methods give quantitative differences, for both functionals the values of interlayer spacings $d_{12}$ and $d_{23}$ decrease after setting greater values of \alfa. The interlayer distances calculated for \alfa=5.02 \AA\, are close to the ones obtained in the previous work based on the GGA approximation\cite{previous} and seem to be the most reasonable values. On the other hand, this configuration proves to be inconsistent with experimental\cite{klusek} n-type doping due to the stretching of C-C bonds, as discussed in the previous section. It is clear that the choice of \alfa\, which allows to preserve the length of the graphene bonds prevents reliable structural optimization. The vertical expansion of gold causes also an increase in doping, which could be equally extrapolated from data in Fig.\ref{dint}. In our opinion, the \alfa\, parameter adapted to graphene needs to be preserved to ensure correct electronic properties of graphene, but the expansion of interlayer spacings in metal should be prevented by keeping its atoms in their positions previously optimized with larger \alfa. In fact, limiting the degrees of freedom of the DFT calculation by fixing the in-plane lattice constant can lead to errors comparable even to those caused by locking the whole system.

\section{Discussion and perspectives}
The theoretical study reported in the previous sections has been focused on the analysis of the model dependent factors that strongly affect the obtained physical properties of graphene on gold. Indeed, the in-plane lattice constant has been treated as a free parameter, although its physical meaning is related to the length of the graphene's bonds. However, we estimated that the adsorption distance hardly depends on \alfa\, suggesting that the interaction with the gold substrate should not considerably stretch the bonds. On the other hand, available experimental data indicate p-type doping of graphene on a gold substrate\cite{klusek}. According to our studies, it means that the C-C bonds do not change their length. In order to obtain correct electronic properties of graphene on Au(111), the lattice constant of graphene should rather be preserved. Effectively, it represents the case where metal atoms with a given coverage were deposited on graphene\cite{interface_xu}. 

It should be noted that in most of our PBE+D2 calculations we have set \alfa=4.89 \AA\, consistently with the LDA-optimized lattice constant of graphene, just to easily compare the results with previous works based on LDA. According to the data presented in Fig.\ref{mapy} (b), setting a real experimental graphene lattice constant ($a=2.46$ \AA) should provide very slight n-type doping. However, the simulations with \alfa=4.932 \AA\, lead to nearly zero doping due to the compensation by relaxation effects. Overall, calculations performed within the DFT-D2 method indicate that graphene is undoped on a clean perfect Au(111) surface meaning that it would be nearly neutral (note that DFT-D2 tends to underestimate the adsorption distance). It could be an explanation why the intercalation of graphene-Ni system by gold atoms leads to almost free-standing graphene. 

We have analyzed the graphene-gold system because contradictory data have been reported in the literature and we demonstrated that these inconsistencies are hardly connected with the choice of the functional. The case of a gold substrate can appear to be complicated, because the transition of n-type to p-type doping is abrupt and occurs in the region of the estimated adsorption distance. However, it seems that the structure and electronic properties of graphene on other metallic substrates, as for example on copper, can be more complex, because the adsorption distance has been reported to change with \alfa\cite{comparative}. We have used the example of gold to unravel the importance of structural details responsible for changes in the Fermi level shift. It is clear that similar methods might be used for characterization of graphene on other metals, especially on Cu and Ag. 

Moreover, our conclusions could be also useful for further studies of molecular layers on the noble metal substrates. Gold surfaces are frequently employed as templates for physisorbed self-assembled monolayers (SAMs)\cite{molekuly1, molekuly2, molekuly3, molekuly4}. Since various choices of geometry and structural optimization methodologies are applied\cite{molekuly5, molekuly6}, our results indicate that similar problems can occur in these systems, especially for the determination of the molecule (assembly)-substrate distance, the values of charge transfer and binding energies.

Finally, we have checked the validity of DFT-D2 method applied to graphene/metal interfaces. We have studied graphene chemisorbed on Ni as well as physisorbed on a Au surface. The results are more reasonable than those obtained within the vdW-DF functional\cite{vanin}, especially the estimation of a graphene-substrate distance is more satisfactory. However, the binding energies seem to be considerably overestimated, which was also reported in studies of the molecules on metals surfaces\cite{aromatic, pentacen}. Of course, a systematic study is required to fully confirm the accuracy of this method. 

We believe that the use of the vdW-DF$^{\mathrm{C}09}$ functional could be equally efficient, since the results in conflict with the experiment reported in Ref.\onlinecite{comparative} might have been caused by the choice of the in-plane lattice constant (see Fig.\ref{mapy}). This configuration seems to describe the samples different from those studied in experiments.
\vskip 1.0in

\begin{acknowledgments}
We thank Z. Klusek for insightful discussions. This work is financially supported by the Polish Ministry of Science and Higher Education in the frame of Grant No. N~N202~204737. One of us (J.S.) acknowledges support from the European Social Fund and Budget of State implemented under the Integrated Regional Operational Program, and from the European Social Fund implemented under the Human Capital Operational Programme (POKL), Project: D-RIM. Part of the numerical calculations reported in this work have been performed at the Interdisciplinary Center for Mathematical and Computational Modelling (ICM) of the University of Warsaw within Grant No. G44\,-\,2. Figure 1 was prepared using the \textsc{xcrysden} program. \cite{kokalj}
\end{acknowledgments}
\end{document}